\documentclass[aps,prd,preprintnumbers,groupedaddress,nofootinbib,showpacs]{revtex4}
\usepackage{graphicx}
\usepackage{xspace}
\usepackage{relsize}

\newcommand{\gev}{\ensuremath{\mathrm{\,Ge\kern -0.1em V}}\xspace}
\newcommand{\mev}{\ensuremath{\mathrm{\,Me\kern -0.1em V}}\xspace}
\newcommand{\kev}{\ensuremath{\mathrm{\,ke\kern -0.1em V}}\xspace}
\newcommand{\tev}{\ensuremath{\mathrm{\,Te\kern -0.1em V}}\xspace}

\newcommand*{\vcenteredhbox}[1]{\begingroup \setbox0=\hbox{#1}\parbox{\wd0}{\box0}\endgroup}

\def\babar{\mbox{\slshape B\kern-0.1em{\smaller A}\kern-0.1em B\kern-0.1em{\smaller A\kern-0.2em R}}}

\def\epem{e^+e^-}
\def\mpmm{\mu^+\mu^-}
\def\lplm{l^+l^-}

\def\CP{\ensuremath{C\!P}\xspace}
\def\epep{e^+e^+}
\def\mpmp{\mu^+\mu^+}
\def\lplp{l^+l^+}

\mathchardef\Upsilon="7107

\begin{document}

\pagestyle{plain}

\title{Search for Light New Physics at $B$ Factories}

\author{Bertrand Echenard,$^1$}
\affiliation{$^1$Department of Physics, California Institute of Technology, Pasadena, CA 91125, USA}
\preprint{CALT-68-2874}
\date{\today}

\begin{abstract}
Many extensions of the Standard Model include the possibility of light new particles, such 
as light Higgs bosons or dark matter candidates. These scenarios can be probed using the large datasets 
collected by $B$ factories, complementing measurements performed at the LHC. This review 
summarizes recent searches for light New Physics conducted by the \babar\ and Belle experiments. 
\end{abstract}


\maketitle
\begin{center} {\it To be published in Advances in High Energy Physics} \end{center}

\section{Introduction}

From supersymmetry to dark matter, many extensions of the Standard Model (SM) include 
the possibility of light New Physics. Thanks to their large luminosities, $B$ factories 
offer an ideal environment to explore these theories. During the last decade, 
the \babar\ Collaboration at PEP-II~\cite{Bib:Babar} 
and the Belle Collaboration at KEKB~\cite{Kurokawa:2001nw,Abashian:2000cg} have respectively collected about 
550 $\rm fb^{-1}$ and more than 1 $\rm ab^{-1}$ of data at several $\Upsilon$ resonances, 
mostly the $\Upsilon(4S)$ resonance (see Table~\ref{Tab:intro}).  These datasets have been 
exploited to explore many aspects of precision physics, including searches for light new 
particles. In the following, we review searches for light Higgs bosons, dark matter candidates, 
hidden sectors, sgoldstinos and Majorana neutrinos. 

\begin{table}[htb]
\begin{center}
\begin{tabular}{l c c c}
\hline
 $\sqrt{s}$  & $\babar$  & Belle & Total \\\hline
$\Upsilon(5S)$   & $-$    &  121 & 121  \\
$\Upsilon(4S)$   & 433    &  711 & 1144  \\
$\Upsilon(3S)$   & 30     &    3 &  33   \\
$\Upsilon(2S)$   & 15     &   25 &  40   \\
$\Upsilon(1S)$   & $-$    &   6  & 6     \\
Off-resonance    & 54     &   94 & 138  \\
\hline
\end{tabular}
\caption{Integrated luminosities ($\rm fb^{-1}$) collected by the $B$ factories at different 
center-of-mass energies. The off-resonance data were collected about $40 \mev$ 
below the $\Upsilon(4S)$ resonance at \babar\, and at a similar offset for the 
$\Upsilon(4S)$ and $\Upsilon(5S)$ resonances in the case of Belle.}
\label{Tab:intro}
\end{center}
\end{table}

\section{Search for light \CP-odd Higgs boson in $\Upsilon$ decays}

A light Higgs boson is predicted by several extensions of the Standard Model, 
such as the Next-to-Minimal Supersymmetric Standard Model (NMSSM). The NMSSM 
Higgs sector contains a total of seven states, three \CP-even, two \CP-odd, 
and two charged Higgs bosons. A \CP-odd Higgs boson ($A^0$) lighter than $2 m_b$ 
can evade present experimental constraints~\cite{Dermisek:2006py}, making it accessible through 
radiative $\Upsilon(nS)\rightarrow \gamma A^0$ decays \cite{Wilczek:1977zn}. The corresponding 
branching fraction could be as large as a few$\times 10^{-4}$, well above the sensitivity of 
$B$ factories~\cite{Dermisek:2006py,Dermisek:2010mg}. 

The Higgs boson decay pattern depends on its mass and couplings, as well as the NMSSM particle 
spectrum. In the absence of light neutralinos, the $A^0$ decays predominantly into a pair of muons  
below $2 m_\tau$, while $\tau^+\tau^-$ and hadronic final states become significant above this 
threshold. The branching fraction $A^0 \rightarrow \chi^0 \bar{\chi}^0$ 
may be dominant if the neutralino ($\chi^0$) is the lightest stable particle with 
$m_{\chi^0} < m_{A^0} /2$~\cite{Shrock:1982kd}. In this case, the neutralino is a natural 
dark matter candidate. 

\babar\ has performed searches for a light \CP-odd Higgs boson in a variety of decay channels. 
These measurements are discussed in the next paragraphs, and the results are summarized in 
Table~\ref{tab:a}. They place stringent constraints on light \CP-odd Higgs models.

\subsection{Search for $\Upsilon(2S,3S) \rightarrow \gamma A^0, A^0 \rightarrow \mu^+\mu^-$}

\begin{figure}[!htb]
\begin{center}
\vcenteredhbox{\includegraphics[width=0.3\textwidth]{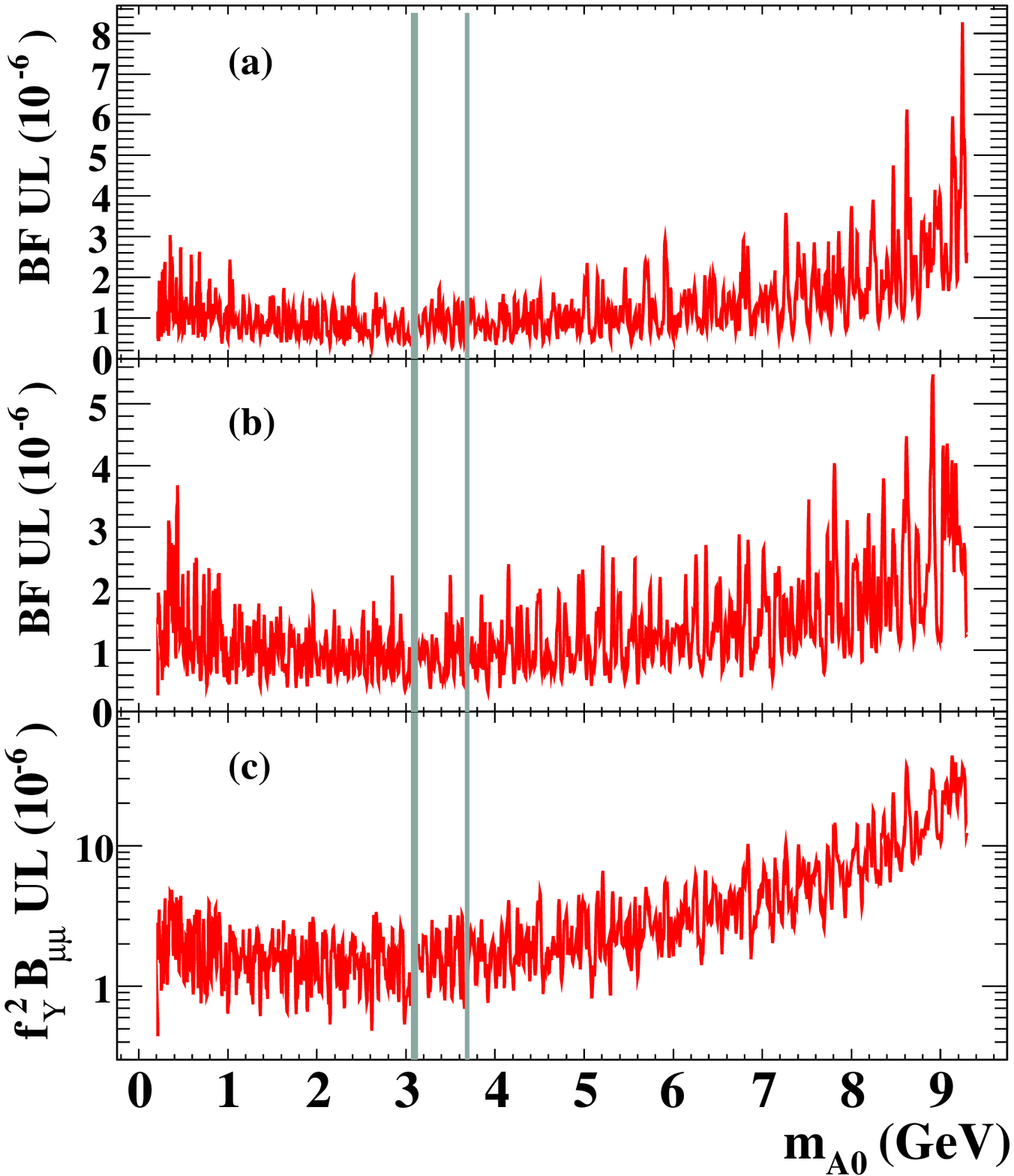} }
\vcenteredhbox{\includegraphics[width=0.65\textwidth]{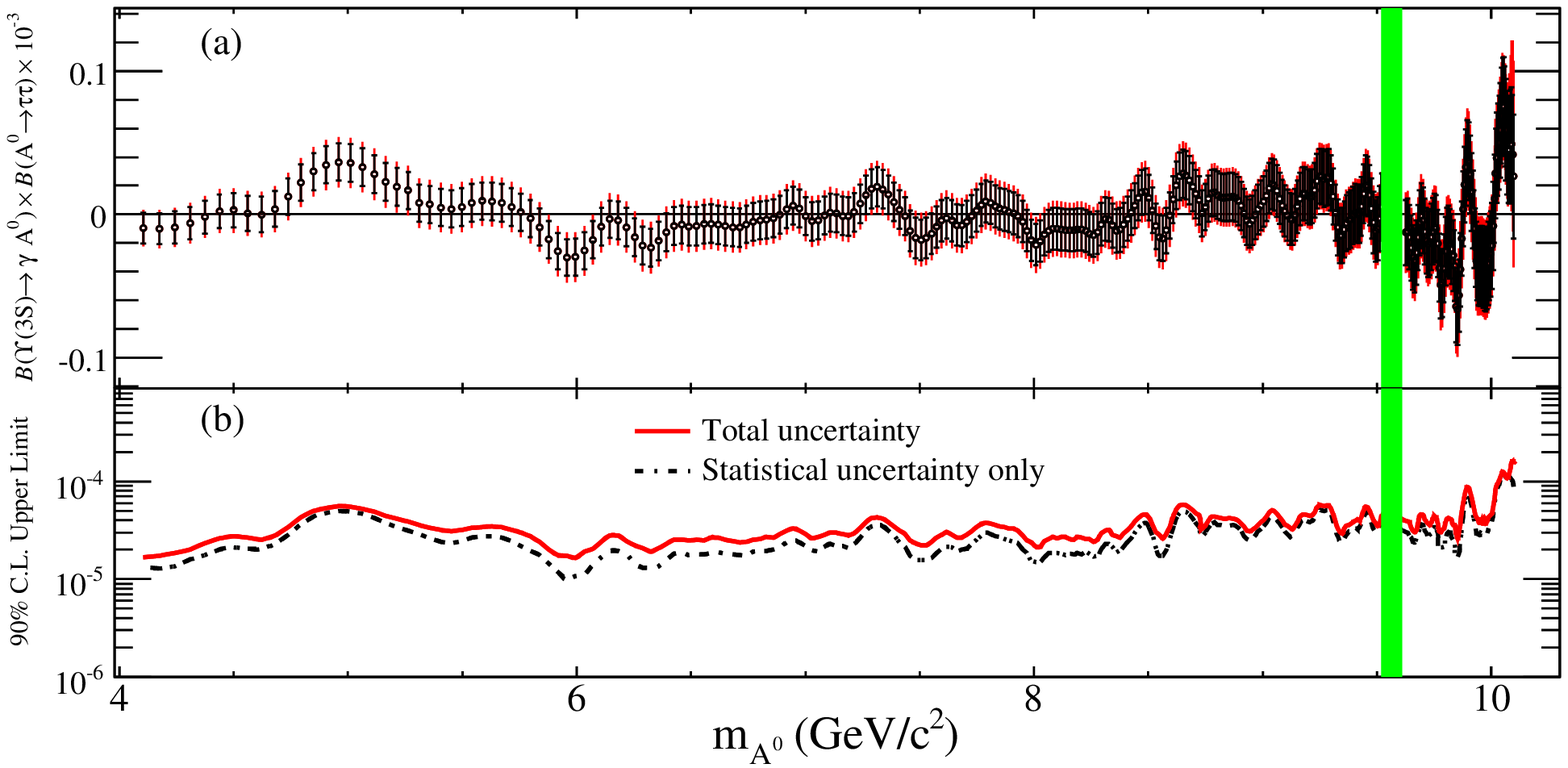} }
\caption{Left: 90\% CL upper limit on the $\Upsilon(3S) \rightarrow \gamma A^0, A^0 \rightarrow \mu^+\mu^-$ (top) 
and $\Upsilon(2S) \rightarrow \gamma A^0, A^0 \rightarrow \mu^+\mu^-$ branching fractions (middle) as a 
function of the $A^0$ mass derived by \babar. The limits on the product $B(A^0 \rightarrow \mu^+\mu^-) f^2_\Upsilon$, are 
also shown (bottom). The $J/\psi$ and $\psi(2S)$ regions (solid bands) are excluded from the search.
Right: Product of branching fraction $\Upsilon(3S) \rightarrow \gamma A^0, A^0 \rightarrow \tau^+\tau^-$ (top) and the 
corresponding 90\% CL upper limit (bottom) as a function of the $A^0$ mass set by \babar. The region corresponding to 
$\chi_{bJ}(2P) \rightarrow \gamma \Upsilon(1S)$ transitions (shaded band) is excluded.}
\label{Fig::Higgs1}
\end{center}
\end{figure}

The $\Upsilon(2S,3S) \rightarrow \gamma A^0, A^0 \rightarrow \mu^+\mu^-$ candidates 
are reconstructed by combining a photon with a pair of oppositely-charged tracks. 
The energy of the photon in the $\Upsilon$ center-of-mass (CM) frame is required to be 
greater than $0.5 \gev$ and one or both tracks must be identified as muons by particle 
identification algorithms. Events containing additional tracks and photons are rejected. 
The $\Upsilon(2,3S)$ candidates are then fit, constraining their CM energies to the 
total beam energy, and imposing a common vertex for the tracks. A series of unbinned 
likelihood fits to the dimuon mass distribution is 
performed to extract the signal. No evidence of $A^0$ is observed; 90\% confidence 
level (CL) limits on the branching fractions are established at the level 
of $(0.26 - 8.3)\times 10^{-6}$ for $0.212 < m_{A^0} < 9.3 \gev$~\cite{Aubert:2009cp}. 
The limits as a function of the $A^0$ mass are shown in Fig.~\ref{Fig::Higgs1}, together 
with limits on the product $B(A^0 \rightarrow \mu^+\mu^-) f^2_\Upsilon$, where 
$f^2_\Upsilon$ denotes the effective coupling of the $b$ quark to the $\Upsilon$ 
meson~\cite{Wilczek:1977zn,Nason:1986tr}. Slightly more stringent constraints have been 
recently derived by the BES-III Collaboration for some $A^0$ mass hypotheses below 
$3 \gev$ using $J/\psi \rightarrow \gamma \mpmm$ decays~\cite{Ablikim:2011es}. 

\subsection{Search for $\Upsilon(3S) \rightarrow \gamma A^0, A^0 \rightarrow \tau^+\tau^-$}

The two taus of the $\Upsilon(3S) \rightarrow \gamma A^0, A^0 \rightarrow \tau^+\tau^-$ 
decays are identified through their leptonic decays, $\tau^+ \rightarrow e^+ \nu_e \bar\nu_\tau$ 
and $\tau^+ \rightarrow \mu^+ \nu_\mu \bar\nu_\tau$. The signal signature consists of 
exactly two oppositely-charged tracks, identified as muons or electrons, and at least 
one photon with an energy greater than $100\mev$ in the CM frame. A set of eight kinematic and angular 
variables are used to further suppress the background, which arises mainly from radiative $\tau$ production and two-photon processes. 
The signal yield is extracted as a function of $m_{A^0}$ by a simultaneous fit to the 
photon energy distribution of the $ee\gamma$, $\mu\mu\gamma$ and $e\mu \gamma$ 
samples. No excess is seen; 90\% CL limits on the branching fraction 
are set at the level of $(1.5-16)\times 10^{-5}$ in the interval 
$4.03 < m_{A^0} < 10.1 \gev$~\cite{Aubert:2009cka}, as shown in Fig~\ref{Fig::Higgs1}.

\subsection{Search for $\Upsilon(2S,3S) \rightarrow \gamma A^0, A^0 \rightarrow \rm \,hadrons$}

The hadronic final states are selected from fully reconstructed the $A^0 \rightarrow \rm \,hadrons$ 
decays. The highest energy photon of the event is identified as the radiative photon from the 
$\Upsilon(nS)$ decay; the $A^0$ candidate is then constructed by adding the four-momenta 
of the remaining particles, constraining the $A^0$ decays products to originate from the interaction 
point to improve the resolution. The signal yield is obtained by fitting the candidate mass spectrum 
in the range $0.3-7.0 \gev$ in steps of $1\mev$. The results are compatible with the null 
hypothesis; limits on the branching fraction are therefore set in the range $(0.1 - 8)\times 10^{-5}$ 
with 90\% confidence level~\cite{Lees:2011wb}.

\subsection{Search for $\Upsilon(2S) \rightarrow \pi^+\pi^- \Upsilon(1S), \Upsilon(1S)\rightarrow \gamma A^0, A^0 \rightarrow  \, \rm invisible$}

\begin{figure}[!htb]
\begin{center}
\includegraphics[width=0.45\textwidth]{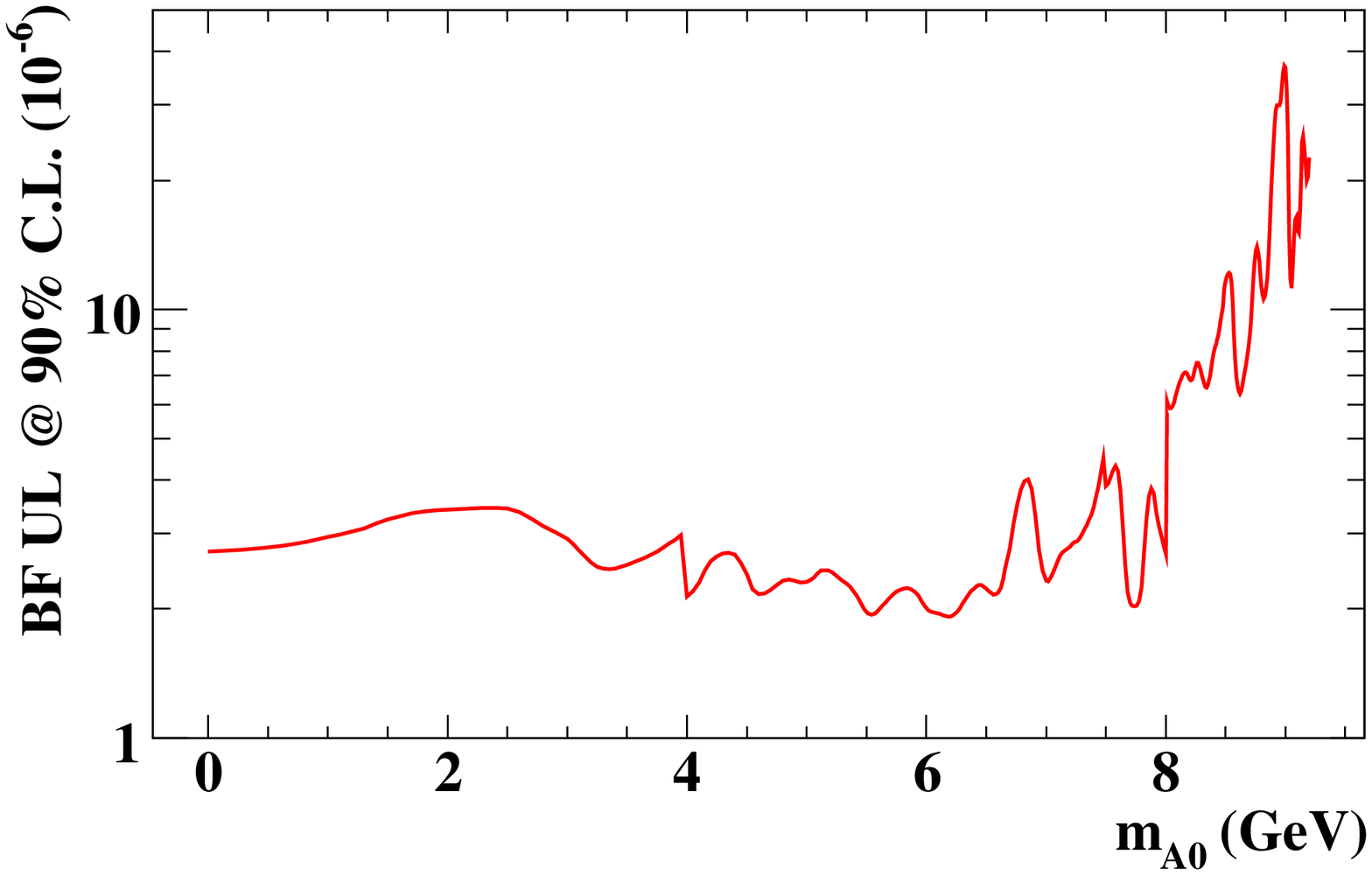}
\includegraphics[width=0.45\textwidth]{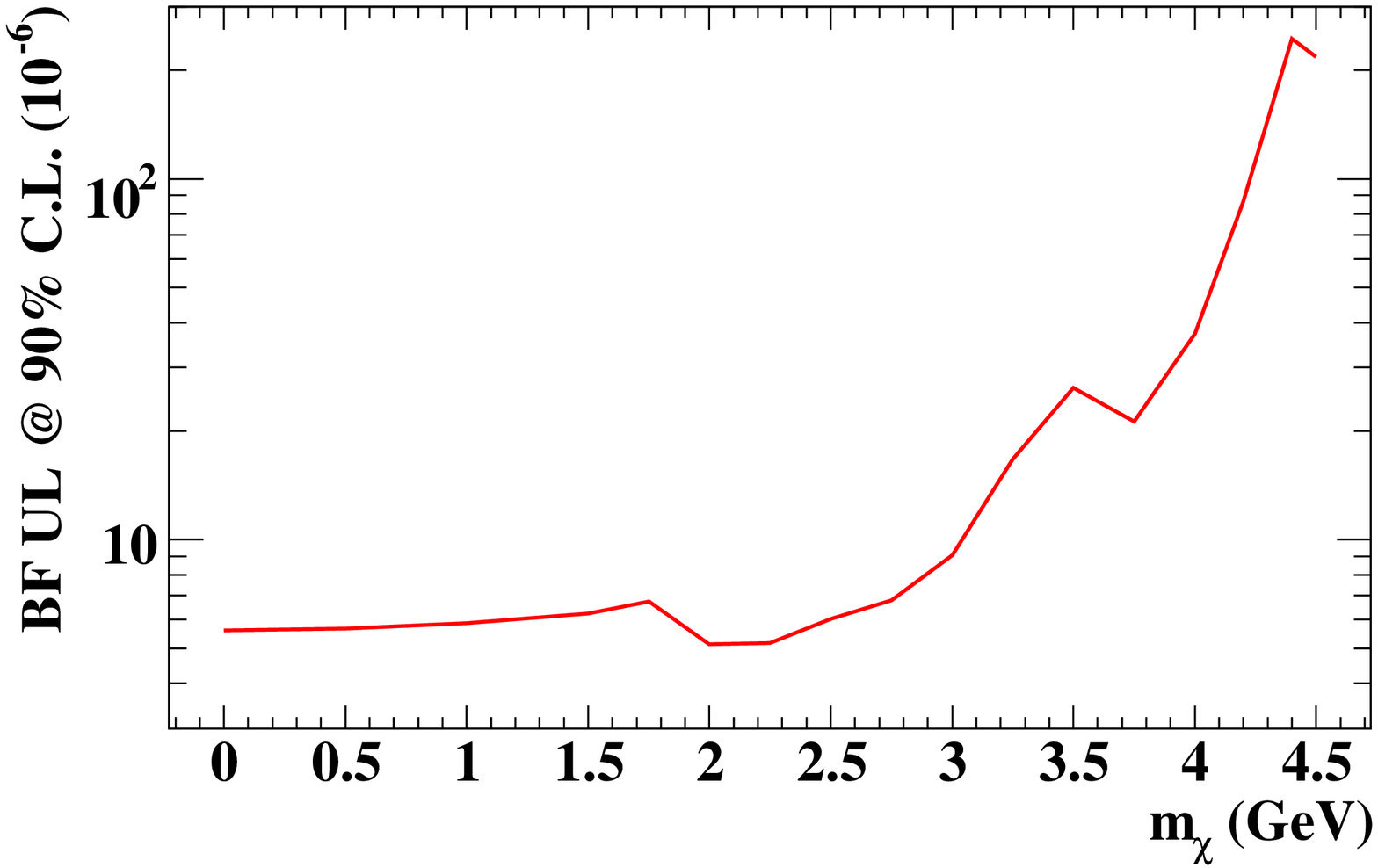}  
\caption{90\% CL upper limits on the $\Upsilon(1S) \rightarrow \gamma  A^0, A^0 \rightarrow \, \rm invisible$ (left) 
and $\Upsilon(1S) \rightarrow \gamma  A^0, A^0 \rightarrow \chi\bar\chi$ (right) branching fractions 
as a function of the $A^0$ and $\chi$ masses set by \babar.}
\label{Fig::Higgs2}
\end{center}
\end{figure}

This final state is characterized by a pair of low-momentum tracks, a single energetic 
photon, and large missing energy and momentum. Additional criteria on the photon 
and the extra neutral energy in the event  are applied to further suppress contributions 
from electron bremsstrahlung, radiative hadronic $\Upsilon(1S)$ decays and 
two-photon processes  in which particles escape undetected. The signal is 
extracted by a series of bidimensional unbinned likelihood fits to the dipion 
recoil mass and the missing mass squared for both two-body decays, 
$\Upsilon(1S) \rightarrow \gamma A^0, A^0 \rightarrow \, \rm invisible$, and non-resonant 
three-body processes, $\Upsilon(1S) \rightarrow \gamma A^0, A^0 \rightarrow \chi \bar\chi$. 
Values of $m_{A^0}$ and $m_{\chi_0}$ are probed over $0 < m_A < 9.2 \gev$ 
and $0 \leq m_{\chi_0} \leq 4.5 \gev$, respectively. No significant signal is found; 90\% CL 
limits $B(\Upsilon(1S) \rightarrow \gamma  A^0, A^0 \rightarrow \, \rm invisible) < (1.9 - 37)\times 10^{-6}$ 
and $B(\Upsilon(1S) \rightarrow \gamma  A^0, A^0 \rightarrow \chi^0\bar{\chi}^0) < (1.9 - 37)\times 10^{-6}$ 
are set~\cite{delAmoSanchez:2010ac}. These limits are displayed in Fig.~\ref{Fig::Higgs2} as a 
function of the $A^0$ and $\chi^0$ masses.

\subsection{Search for $\Upsilon(3S)\rightarrow \gamma A^0, A^0 \rightarrow \, \rm invisible$}

The $\Upsilon(3S) \rightarrow  \gamma A^0, A^0 \rightarrow \, \rm invisible$ decays are also 
selected from events containing a single energetic photon with no additional 
activity in the detector. The background arises mainly from 
$\epem \rightarrow \gamma \gamma$, radiative Bhabha, and two-photon 
fusion events. The $A^0$ yield is extracted by a series of unbinned likelihood fits 
to the photon energy distribution for $0 < m_{A^0} < 7.8 \gev$. No excess is seen, 
and limits on the branching fraction at the level of $(0.7 - 31)\times 10^{-6}$ are 
derived with 90\% confidence level~\cite{Aubert:2009ad}.

\begin{table}
\begin{tabular}{lrr}
\hline
Mode  & Mass range ($\gev$)& BF upper limit (90\% CL)\\
\hline
$\Upsilon(2S,3S) \rightarrow \gamma A^0, A^0 \rightarrow \mu^+\mu^-$        & $0.21 < m_A < 9.3$   & $(0.3 - 8.3)\times 10^{-6}$\\
$\Upsilon(3S)    \rightarrow \gamma A^0, A^0 \rightarrow \tau^+\tau^-$      & $4.0  < m_A < 10.1$  & $(1.5 - 16)\times 10^{-5}$\\
$\Upsilon(2S,3S) \rightarrow \gamma A^0, A^0 \rightarrow \rm \,hadrons$     & $0.3 < m_A < 7.0$    & $(0.1 - 8)\times 10^{-5}$\\
$\Upsilon(1S)    \rightarrow \gamma A^0, A^0 \rightarrow \chi\bar\chi$      & $m_\chi< 4.5 \gev$   & $(0.5 - 24)\times 10^{-5}$\\
$\Upsilon(1S)    \rightarrow \gamma A^0, A^0 \rightarrow \, \rm invisible$  & $m_A < 9.2 \gev$     & $(1.9 - 37)\times 10^{-6}$\\
$\Upsilon(3S)    \rightarrow \gamma A^0, A^0 \rightarrow \, \rm invisible$  & $m_A < 9.2 \gev$     & $(0.7 - 31)\times 10^{-6}$\\
\hline
\end{tabular}
\caption{Results of light Higgs boson searches performed by the \babar\ Collaboration.}
\label{tab:a}
\end{table}

\section{Search for dark matter in invisible $\Upsilon(1S)$ decays}

In a minimal model, a single dark matter particle ($\chi$) is added to 
the SM content, together with a new boson mediating SM-dark matter 
interactions~\cite{McElrath:2005bp,Yeghiyan:2009xc,Fayet:2009tv}. A 
light mediator could be produced in $b\bar{b}$ annihilation and decay into a 
$\chi \bar{\chi}$ pair, contributing to the invisible width of $\Upsilon$ mesons. 
In the SM, invisible $\Upsilon(1S)$ decays proceed via the production of 
a $\nu \bar{\nu}$ pair with a branching fraction 
$B(\Upsilon(1S) \rightarrow \nu \bar{\nu}) \sim (1 \times 10^{-5})$~\cite{Chang:1997tq}, 
well below the current experimental sensitivity. The rate 
$\Upsilon(1S) \rightarrow \chi \bar{\chi}$ is predicted to be larger 
by one or two orders of magnitude than that of $\Upsilon(1S) \rightarrow \nu \bar{\nu}$, 
assuming no flavor changing currents~\cite{McElrath:2007sa}.

A search for dark matter in invisible $\Upsilon(1S)$ decays has been 
performed by $\babar$ using a sample of $122\times 10^6$ $\Upsilon(3S)$ 
mesons~\cite{Aubert:2009ae}. The $\Upsilon(1S)$ mesons are selected 
by reconstructing the $\Upsilon(3S) \rightarrow \pi^+\pi^- \Upsilon(1S)$ transitions. 
The dipion recoil mass peaks at the $\Upsilon(1S)$ for signal events, while 
backgrounds are broadly distributed. This strategy is common to 
several analyses, and provides a very clean $\Upsilon(1S)$ sample. 

The event topology consists of exactly two oppositely-charged tracks 
without any additional activity. The selection is performed using a 
multivariate classifier based on variables describing the pions, the 
neutral energy deposited in the calorimeters and the multiplicity 
of $K^0_L$ candidates. 

\begin{figure}[htb]
\begin{center}
\includegraphics[width=0.4\textheight]{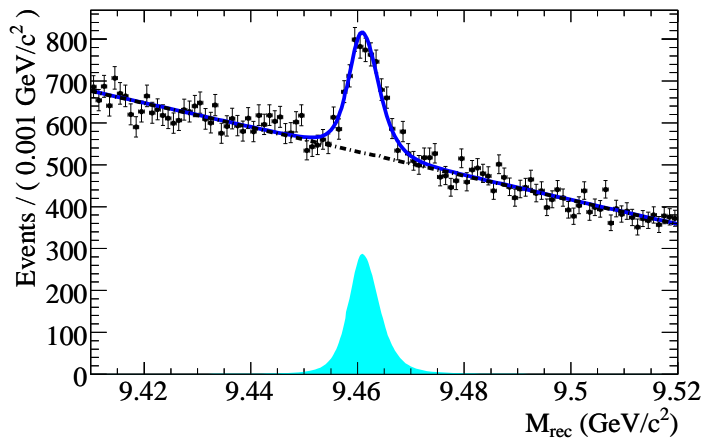} 
\end{center}
\caption{The distribution of the dipion recoil mass ($M_{rec}$) for \babar\ data, together with the 
result of the maximum likelihood fit (full line). The non-resonant background (dashed line) 
and the sum of the signal and the peaking background (solid filled) are also shown.}
\label{Fig::inv1}
\end{figure}

The distribution of the resulting dipion recoil mass (Fig.~\ref{Fig::inv1}) 
shows a clear peak corresponding to $\Upsilon(1S)$ mesons on top 
of a non-resonant component. In addition to signal events, a background 
from $\Upsilon(1S)$ decays in which the decay products escape 
undetected, is also present. This component, kinematically 
indistinguishable from the signal, is evaluated using Monte 
Carlo simulations. 

The sum of signal and peaking background yields is first extracted by 
an extended maximum likelihood fit to the dipion recoil mass. After 
subtracting the peaking background, a signal yield of 
$-118 \pm 105 \pm 124$ is measured, where the first uncertainty is statistical 
and the second systematic. No evidence for $\Upsilon(1S) \rightarrow \, \rm invisible$ 
decays is observed and a 90\% confidence level Bayesian upper limit on its 
branching fraction is set at $3.0 \times 10^{-4}$ using a prior flat in branching 
fraction. This result improves the best previous measurement~\cite{Tajima:2006nc} 
by nearly an order of magnitude, and sets stringent constraints on minimal light 
dark matter models.

\section{Search for dark matter and hidden sectors}

A new class of dark matter model has recently been proposed, following observation 
from satellite and ground-based experiments. These models introduce a 
new hidden sector with WIMP-like fermionic dark matter particles charged 
under a new Abelian gauge group~\cite{Fayet,Pospelov:2007mp,ArkaniHamed}. 
The corresponding gauge boson, dubbed a hidden photon ($A'$), is constrained 
to have a mass at the $\gev$ scale to explain the electron/positron excess 
observed by PAMELA~\cite{Adriani:2010ib} and FERMI~\cite{FermiLAT:2011ab}, 
without a comparable anti-proton signal. The hidden photon couples to the 
SM photon through kinetic mixing with a mixing strength $\epsilon$, connecting 
the hidden sector to SM particles~\cite{Holdom}.

The Higgs mechanism generates the hidden boson masses, adding hidden 
Higgs bosons $(h')$ to the theory. A minimal model includes a single hidden 
photon and a Higgs boson~\cite{Batell:2009yf}. Additional gauge and Higgs 
bosons are considered in more complex variations~\cite{Essig:2009nc,Baumgart:2009tn}.

\subsection{Search for a hidden photon}

\begin{figure}[hbt]
\begin{center}
\includegraphics[width=0.4\textwidth]{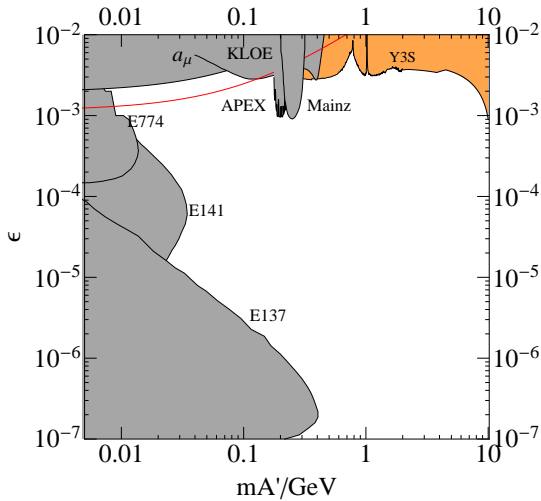}
\caption{Constraints on the mixing strength, $\epsilon$, as a function of the 
hidden photon mass derived from searches in $\Upsilon(2S,3S)$ decays at \babar\ 
(orange shading) and from other experiments (gray shading). 
The red line shows the value of the coupling required to 
explain the discrepancy between the calculated and measured anomalous magnetic 
moment of the muon~\protect\cite{Pospelov:2008zw}.}
\label{Fig::hp}
\end{center}
\end{figure}

Hidden photons can be readily formed in $\epem \rightarrow \gamma A'$ 
interactions, and be reconstructed via their leptonic decays as resonances 
in the $\epem \rightarrow \gamma l^+l^-$ ($l=e,\mu$) spectrum. 
This signature is similar to that of light \CP-odd Higgs production in 
$\epem \rightarrow \Upsilon(2S,3S) \rightarrow \gamma \mu^+\mu^-$~\cite{Aubert:2009cp}, 
and searches for this channel have therefore been reinterpreted as constraints on hidden photon 
production~\cite{Bjorken:2009mm}. The limits are shown in Fig.~\ref{Fig::hp}, together 
with bounds derived from measurements of $\phi \rightarrow \eta A', A' \rightarrow 
\epem$ decays at KLOE~\cite{Giovannella:2011nh}, and searches for direct 
production in fixed-target experiments~\cite{Bjorken:2009mm,Merkel:2011ze,Abrahamyan:2011gv}. A 
hidden photon could also contribute to the muon anomalous magnetic moment, and 
constraints from this measurement are also shown. Values of the mixing strength 
down to $10^{-3} - 10^{-2}$ are probed for $0.212 < m_{A'} < 9.3 \gev$. 

Other measurements could be reinterpreted as bounds on hidden photon production, 
such as searches for peaks in $\epem \rightarrow \gamma \tau^+\tau^-$  
events~\cite{Aubert:2009cka} or inclusive $\epem \rightarrow \gamma ~\rm{hadrons}$ 
production~\cite{Lees:2011wb}. Invisible decays, occurring if hidden bosons decay 
to long-lived states, could also be detected as a mono-energetic photon peaks 
in $\epem \rightarrow \gamma +\rm{invisible}$ events~\cite{Aubert:2009ad}.

\subsection{Search for hidden sector bosons}

Non-Abelian extensions of hidden sectors introduce additional hidden gauge bosons, generically 
denoted $W',W'',...$. The phenomenology depends on the precise structure of the model, but heavy 
hidden bosons decay to lighter states if kinematically accessible, while the lightest bosons are 
metastable and decay to SM fermions via their mixing with the hidden photon~\cite{Essig:2009nc,Baumgart:2009tn}. 
Non-Abelian hidden sectors could also accommodate inelastic dark matter~\cite{TuckerSmith:2001hy} if the 
mass spectrum contains nearly degenerated states.

\babar\ has performed a search for di-boson production in 
$\epem \rightarrow A'^* \rightarrow W' W'', W'\rightarrow l^+l^-, W''\rightarrow l^+l^-$ ($l=e,\,u$) 
events, where the bosons are reconstructed via their decays into lepton pairs~\cite{arXiv:0908.2821}. 
The study has been performed in the context of inelastic dark matter models, searching for two bosons 
with similar masses.

The signal signature consists of two narrow dileptonic resonances with similar masses carrying the 
full beam energy. This topology is quite unique; the only backgrounds arise from QED processes. The 
signal is extracted as a function of the average dileptonic mass in the range $0.24 - 5.3 \gev$. No 
significant signal is found; limits on the $\epem \rightarrow A'^* \rightarrow W' W''$ cross-section 
are derived. The results are translated into limits on the product $\alpha_D \epsilon^2$, where 
$\alpha_D = g_D^2/4\pi$ and $g_D$ is the hidden sector gauge coupling constant. Values down to 
$10^{-10}$ are probed, assuming nearly degenerate bosons.

\subsection{Search for a hidden Higgs boson}

The Higgsstrahlung process, $\epem \rightarrow A' h', h' \rightarrow A' A'$, 
offers another gateway to hidden sectors. This process is one 
of the few suppressed by only a single power of the mixing strength, and the 
background is expected to be almost negligible. The event topology is driven 
by the boson masses. While Higgs bosons heavier than two hidden photons 
decay promptly, they become metastable below this threshold 
and either produce displaced decays or escape undetected.

A search for hidden Higgs boson in Higgsstrahlung production in the prompt decay 
regime has been conducted at \babar, based on a data sample of 521 fb$^{-1}$ ~\cite{Lees:2012ra}. 
The measurement is performed in the range $0.8 < m_{h'} < 10.0 \gev$ and 
$0.25 < m_{A'} < 3.0 \gev$, with the constraint $m_{h'} > 2 m_{A'}$. 
The signal is either fully reconstructed into lepton or pion pairs (exclusive mode), or partially 
reconstructed (inclusive mode). The exclusive modes contain of six tracks, forming three 
hidden photon candidates with equal masses and a total invariant mass close to the 
$\epem$ CM energy. The six pion final state has a significantly larger background than the other exclusive modes 
and is excluded from the search. Only two of the three hidden photons are reconstructed as dileptonic 
resonances for the inclusive modes. The remaining hidden photon, assigned to the 
recoiling system, must have a mass compatible with the Higgsstrahlung hypothesis. 

\begin{figure}[!htb]
\begin{center}
\includegraphics[width=0.5\textwidth]{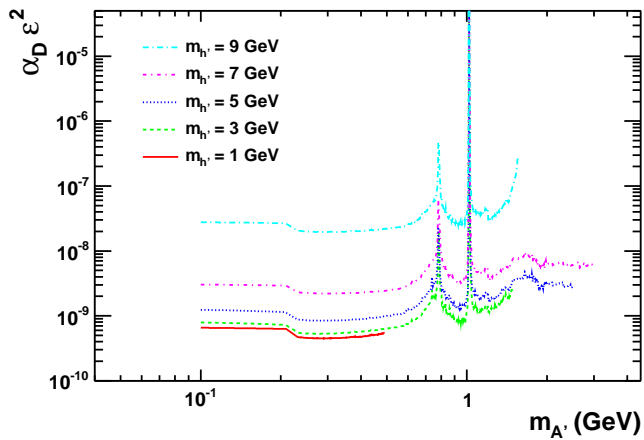} 
\caption{The 90\% confidence level upper limit derived by \babar\ on the product $\alpha_D \epsilon^2$ as a function 
of the hidden photon mass ($m_{A'}$) for selected values of hidden Higgs boson masses ($m_{h'}$).}
\label{Fig::higgs2}
\end{center}
\end{figure}

No significant signal is observed, and upper limits on the $e^+ e^- \rightarrow  A' h', h' \rightarrow A' A'$ 
cross section are set as a function of the hidden Higgs and hidden photon masses. These bounds 
are finally converted into limits on the product $\alpha_D \epsilon^2$. The results are displayed 
in Fig.~\ref{Fig::higgs2}. Values down to $10^{-10} - 10^{-8}$ are excluded for a substantial fraction 
of the parameter space probed. These limits are translated into constraints on the mixing strength 
in the range $10^{-4} - 10^{-3}$, assuming $\alpha_D = \alpha \simeq 1/137$.

\section{Search for dimuon decays of pseudoscalar sgoldstinos}

The observation of three $\Sigma^+ \rightarrow p \mu^+\mu^-$ events 
with a dimuon invariant mass clustered around $214 \mev$ by 
the HyperCP Collaboration \cite{Park:2005eka} triggered much discussion 
about the possibility of a new light state $X$ produced in  
$\Sigma^+ \rightarrow X, X \rightarrow \mu^+\mu^-$ decays. Speculations 
about the nature of this state included a pseudoscalar sgoldstino~\cite{Gorbunov:2005nu}, a 
hidden sector photon~\cite{Chen:2007uv,Pospelov:2008zw} or a light Higgs boson~\cite{He:2006fr}. 
Subsequent measurements in $e^+e^-$ collisions \cite{Aubert:2009cp,Merkel:2011ze} and 
$p \bar{p}$ interactions \cite{Abazov:2009yi} excluded the light Higgs boson and 
hidden photon hypotheses.

\begin{figure}[htb]
\begin{center}
\includegraphics[width=0.45\textwidth]{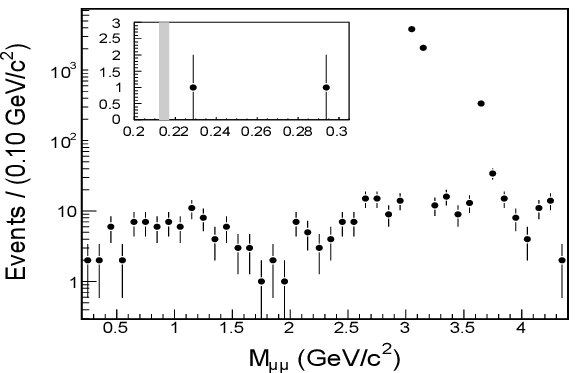} 
\includegraphics[width=0.45\textwidth]{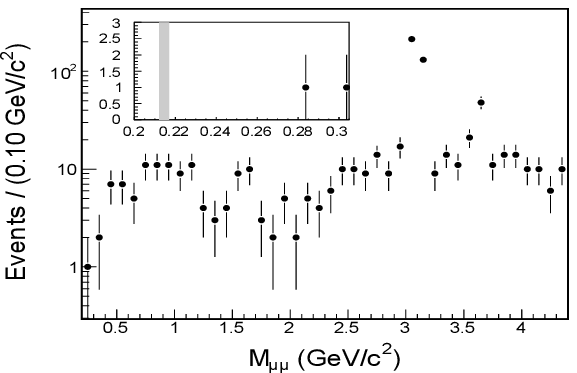} 
\end{center}
\caption{Distribution of the dimuon invariant mass for $B \rightarrow K^{*0}\mu^+\mu^-$ (left) and 
$B \rightarrow \rho^0 \mu^+\mu^-$ (right) signal events for Belle data. The signal regions are shown as dashed bands 
in the inserts.} 
\label{Fig::sgold}
\end{figure}

The Belle Collaboration performed a search for this state in 
$B \rightarrow K^{*0}X, K^{*0} \rightarrow K^+\pi^-, X \rightarrow \mu^+\mu^-$ ($B_{K^{*0}X}$) and 
$B \rightarrow \rho^0 X, \rho^0 \rightarrow \pi^+\pi^-, X \rightarrow \mu^+\mu^-$ ($B_{\rho^0X}$) 
decays using a sample of 675 millions $B^0 \bar{B}^0$ pairs~\cite{Hyun:2010an}. The 
$B$ mesons are reconstructed by combining two well-identified muons with a $K^+\pi^-$ or 
$\pi^+\pi^-$ pair. The signal candidates are identified using the beam-energy constrained mass 
$M_{bc} = \sqrt{E_{beam}^2-p_B}$ and the energy difference $\Delta E = E_{beam}-E_B$, where $E_{beam}$ 
denotes the beam energy and $p_B$ ($E_b$) the momentum (energy) of the $B$ candidate in the $\epem$ 
center-of-mass frame. Signal $B_{K^{*0}X}$ ($B_{\rho^0X}$) events are selected in the region 
$5.27<M_{bc}<5.29 \gev$ and $-0.03 < \Delta E < 0.04 \gev$ ($-0.04 < \Delta E < 0.04 \gev$). The 
corresponding dimuon mass distributions are shown in Fig.~\ref{Fig::sgold}. No events are observed 
in the signal region.

Lacking evidence for a pseudoscalar dimuon resonance at $214.3 \mev$, 90\% C.L. upper limits on the 
branching fraction $B \rightarrow K^{*0}X$ and $B \rightarrow \rho^0X$ are set at the 
level of $2.26 \times 10^{-8}$ and $1.73 \times 10^{-8}$, respectively. They rule out models II and III 
of the sgoldstino interpretation of the HyperCP observation \cite{Demidov:2006pt}.

\section{Search for lepton number violation and a Majorana neutrino}

Lepton number is conserved in low-energy processes in the SM 
model, but can be violated in a number of New Physics scenarios, 
such as models containing Majorana neutrinos. In this case, the 
neutrino is its own antiparticle, and reactions changing the lepton 
number by two units become possible. The most sensitive 
searches have so far been based on neutrinoless nuclear 
double beta decays $0\nu \beta \beta$ \cite{GomezCadenas:2011it}, but the nuclear 
environment complicates the extraction of the neutrino 
mass scale. Processes involving meson decays, such as 
$B \rightarrow h^- \lplp$ ($h=\pi,K,D$ and $l=e,\mu$), have been 
proposed as a possible alternative. The 
presence of a Majorana neutrino could mediate such a reaction, and 
would appear as an enhanced peak in the mass spectrum of the hadron 
and one of the leptons~\cite{Zhang:2010um,Atre:2009rg}. 

\babar\ has performed a search for the lepton number violating 
$B^+ \rightarrow h^- \lplp$ decays with $h=\pi,K$ and $l=e,\mu$, 
based on a sample of $471 \pm 3 $ millions $B\bar{B}$ pairs~\cite{BABAR:2012aa}. 
The $B$ meson candidates are reconstructed by combining a 
hadron with a pair of tracks identified as leptons from particle 
identification algorithms. The background is 
suppressed through boosted decision trees (BDTs) using variables 
describing the event shape and the $B$ meson candidate. The number 
of $B \rightarrow h^- \lplm$ events is extracted by a multidimensional 
likelihood fit of the BDTs response and the beam-energy constrained mass 
$m_{ES} = \sqrt{s/4 -p_B^{*2}}$, where $p_B^{*}$ is the momentum 
of the $\epem$ CM frame. No evidence for such decays is observed, 
and limits on the corresponding branching fractions are set. 

A similar analysis has been conducted by Belle in $B^+ \rightarrow D^- \epep$, 
$B^+ \rightarrow D^- e^+\mu^+$ and $B^+ \rightarrow D^- \mpmp$  
decays, followed by a subsequent $D^- \rightarrow K^+ \pi^- \pi^-$ decay~\cite{Seon:2011ni}. The 
analysis is based on a data sample of 772 million $B\bar{B}$ pairs 
collected at the $\Upsilon(4S)$ resonance. The $B$ candidates are identified
using the energy difference, $\Delta E = E_B - E_{beam}$ and $m_{ES}$. 
The signal region is defined as $ 5.27 < m_{ES} < 5.29 \gev$ and 
$-0.055 (-0.035) < \Delta E < 0.035 \gev$ for the $e^+e^+$ and 
$e^+\mu^+$ ($\mu^+\mu^+$) final states. No signal events are observed 
and 90\% CL limits on the branching fractions are set, assuming uniform 
three-body phase space distributions for $B^+ \rightarrow D^- l^+ l'^+$ 
decays.

\begin{figure}[htb]
\begin{center}
\includegraphics[width=0.6\textwidth]{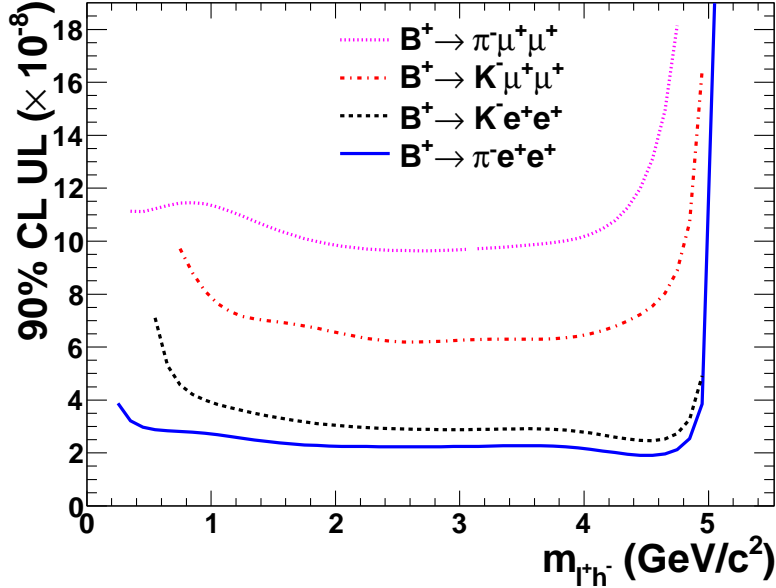} 
\caption{90\% CL upper limits (UL) on the branching fraction $B(B^+ \rightarrow h^- \lplp)$ as a function of the 
$h^- l^+$ mass ($m_{h^- l^+}$) derived by \babar.}
\label{Fig::LNV}
\end{center}
\end{figure}

The results are reported in Table~\ref{Tab:LNFV}, and the limits on 
$B^+ \rightarrow h^- \lplp$ decays as a function of the Majorana 
neutrino mass $m_\nu = m_{l^+h^-}$ are also displayed in Fig.~\ref{Fig::LNV}. 
The sensitivities of the $B \rightarrow h^- \mpmp$ channels is similar to that of recent 
measurements from LHCb in the same channel~\cite{Aaij:2011ex,Aaij:2012zr}, 
and are an order of magnitude more stringent than the previous results for 
$B^+ \rightarrow h^- \epep$ decays \cite{Edwards:2002kq}.

\begin{table}
\begin{center}
\begin{tabular}{| c c || c c|}
\hline
Mode & BF UL ($10^{-8}$) & mode & BF UL  ($10^{-6}$) \\\hline
$B^+ \rightarrow \pi^- e^+e^+$        & 3.0     & $B^+ \rightarrow D^- \epep$       & 2.6 \\
$B^+ \rightarrow K^- e^+e^+$          & 2.3     & $B^+ \rightarrow D^- e^+\mu^+$    & 1.8\\
$B^+ \rightarrow \pi^- \mu^+ \mu^+$   & 10.7    & $B^+ \rightarrow D^- \mpmp$       & 1.0\\
$B^+ \rightarrow K^- \mu^+ \mu^+$     & 6.7     & \multicolumn{2}{c|}{}\\
\hline
\end{tabular}
\caption{Results of lepton number violation searches performed by the \babar\ (left column) and Belle Collaborations 
(right column) . The limits are given at 90\% confidence level.}
\label{Tab:LNFV}
\end{center}
\end{table}  

\section{Conclusion}

$B$ factories have proved to be versatile machines, ideally suited to search for light 
New Physics over a wide range of processes. The next generation of flavor factories 
are expected to improve the sensitivity of these searches by one to two orders of magnitude, further 
constraining the parameter space of these theories. Many more results are to come 
in the near future, and will hopefully contribute to elucidate the nature of physics 
beyond the Standard Model.

\section*{Acknowledgments}

I would like to thank David Hitlin for his comments on this manuscript, Rouven Essig for useful theoretical 
discussions and Matthew Graham for discussing the constraints on dark photon production and providing the 
corresponding figure. BE is supported by the Department of Energy, under grant DE-FG02-92ER40701.

\end{document}